# An integrated low phase noise radiation-pressure-driven optomechanical oscillator chipset

Xingsheng Luan[1,*], Yongjun Huang[1,2,*], Ying Li[1], James F. McMillan[1], Jiangjun Zheng[1], Shu-Wei Huang[1], Pin-Chun Hsieh[1], Tingyi Gu[1], Di Wang[1], Archita Hati[3], David A. Howe[3], Guangjun Wen[2], Mingbin Yu[4], Guoqiang Lo[4], Dim-Lee Kwong[4], and Chee Wei Wong[1,*]

[1]*Optical Nanostructures Laboratory, Columbia University, New York, NY 10027, USA*

[2]*Key Laboratory of Broadband Optical Fiber Transmission & Communication Networks, School of Communication and Information Engineering*

*University of Electronic Science and Technology of China, Chengdu, 611731, China*

[3]*National Institute of Standards and Technology, Boulder, CO 80303, USA*

[4]*The Institute of Microelectronics, 11 Science Park Road, Singapore 117685, Singapore*

*Author e-mail address: xl2354@columbia.edu, yh2663@columbia.edu; cww2104@columbia.edu

**High-quality frequency references are the cornerstones in position, navigation and timing applications of both scientific and commercial domains. Optomechanical oscillators, with direct coupling to continuous-wave light and non-material-limited $f \times Q$ product, are long regarded as a potential platform for frequency reference in radio-frequency-photonic architectures. However, one major challenge is the compatibility with standard CMOS fabrication processes while maintaining optomechanical high quality performance. Here we demonstrate the monolithic integration of photonic crystal optomechanical oscillators and on-chip high speed Ge detectors based on the silicon CMOS platform. With the generation of both high harmonics (up to 59th order) and subharmonics (down to 1/4), our chipset provides multiple frequency tones for applications in both frequency multipliers and dividers. The phase noise is measured down to -125 dBc/Hz at 10 kHz offset at ~ 400 µW dropped-in powers, one of the lowest noise optomechanical oscillators to date and in room-temperature and atmospheric non-vacuum operating conditions. These characteristics enable optomechanical oscillators as a frequency reference platform for radio-frequency-photonic information processing.**

Till now the most widely used commercial frequency references are based on quartz crystal oscillators which, after more than eight decades of development, have achieved remarkable low phase noise performance. However, due to the incompatibility with standard CMOS processes, the quartz oscillator is also well-known as one of the last electronic components that have yet to yield to silicon integration. Thereby there is a strong motivation to develop high-quality silicon oscillators as early as 1980s. Compared with quartz oscillators, silicon oscillators have smaller size, lower cost and power consumption and most importantly, can potentially be fabricated by standard CMOS processes with ease of integration to silicon electronic circuits. Recently the emergence of optomechanical oscillators (OMO)[1] as a photonic clock provides an alternative approach towards stable chip-scale radio frequency(RF) references[2,3].

In optomechanical oscillators, the mechanical resonator and optical cavity are designed on the same device to maximize the optomechanical coupling. With carefully-tuned high quality factor ($Q$) and tight sub-wavelength confinement of selected optical cavities, large radiation pressure forces[4] can be possible, modifying the motion of micro/nano-mechanical resonators[5–7]. When the input (drive) optical power exceeds the intrinsic mechanical damping losses, the mechanical resonator becomes a self-sustained oscillator[8,9] with quantum backaction limited linewidth[10]. The optical amplified periodic motion of the mechanical resonator perturbs the optical cavity resonance, transducing the mechanical motion into the intracavity optical field. Such periodic modulation can be optically read out by measuring the optical transmission from the cavity, thus making an on-chip photonic-based RFreference[2,3,11,12]. Unlike quartz crystal oscillators, the optomechanical oscillator performance is not limited by the $f \times Q$ product of the material[13] and their linewidth is limited only by quantum dynamical backaction[10] and phase noise of drive laser[14]. Recent efforts in improving the performance of mesoscopic optomechanical oscillators include development of several novel optomechanical frequency stabilization techniques[15,16] and realization of different optomechanical cavity configurations[8,12,17,18].

However, for chip-scale operations with integrated electronics, CMOS-compatibility remains as a challenge for further applications of optomechanical oscillator. For example, a fully monolithically-integrated OMO with on-board detector and electronics can provide a portable frequency reference, potentially lower close-to-carrier (e.g. 1 kHz or less) phase noise, and allows as much RF power into the detectors as possible for signal processing. Although integration with Ge detectors was recently examined with ring oscillators[19], the optomechanical transduction for ring optomechanical oscillators are about one to two orders-of-magnitude weaker than photonic crystal optomechanical cavities[20–23], resulting in high pump power operation[11], weak signals that demand vacuum operating requirements, and/or the auxiliary of an electric driving force[19] which could introduce extra noise. Here we demonstrated the monolithic integration of photonic crystal optomechanical oscillators with on-chip Ge detectors, with large zero-point optomechanical coupling strength ~ 800 kHz and a resulting high-harmonic up to ~ 7 GHz. Furthermore, we observed novel fractional sub-harmonics generation and demonstrated injection locking of fundamental mode and high-order harmonics simultaneously in our CMOS-compatible chipset. Consequently we report for the first time the optomechanical chipset with low phase noise down to -125 dBc/Hz at 10 kHz offset, one of the lowest noise optomechanical oscillators to date and with ambient (atmosphere, non-vacuum, and room temperature) operations[2,3,11,24]. The integration of the photonic crystal OMO with active optoelectronics is non-trivial, involving high-performance nanomembrane optical cavities with 120 nm critical dimensions next to Si-Ge molecular beam epitaxial growth, junction electronics, and optimized optical components across multilayer planarization and processing.

Figure 1a shows the fully integrated optomechanical cavity oscillator and on-chip Ge detector chipset, with the optical waveguide path denoted in green (detailed description in Supplementary Information I). Laser is first coupled from free-space lenses into a low-loss inverse oxide coupler at the chip facet (left side of Figure 1a), propagating then from the oxide coupler into a silicon waveguide. Before entering the PhC cavity, another inverse taper is also designed to ensure high efficiency tunneling into the photonic crystal waveguide and the cavity center. The transmitted light is split equally into two paths: one into the integrated Ge detector and the other coupled out from the inverse oxide coupler to off-chip lenses for external test diagnostics. Such test design allows us to monitor the comparative signal from the integrated Ge detector and external detector simultaneously. The optomechanical oscillator examined is a slot-type photonic crystal (PhC) cavity consisting of two (16.0 μm × 5.5 μm) air-bridged photonic crystal slabs separated by a narrow 120 nm air slot as shown in Figure 1b to 1d. Slot-type PhC have been studied previously via electron-beam lithography[25–27], with a slot-guided cavity mode of optical quality factor ~$10^6$, These optical modes couple to the fundamental mechanical mode with a vibration frequency ~100 MHz and quality factor ~ $10^3$ in room temperature and atmosphere[12,28]. The tight photon confinement in optomechanical photonic crystals[12,21,28] allows large radiation pressure effects, especially in our sub-wavelength slot cavities, which has a strong vacuum optomechanical coupling strength ~ 2.5 MHz in modeling[21] and ~ 800 kHz in experiment[12]. The localized slot guided mode is formed by first introducing a line-defect through removing and shifting a central line of air holes in a periodic optical lattice (see Figure 1b to 1d). The line-defect width, defined as the distance between the center of adjacent holes, is set to be $1.2 \times \sqrt{3}a$ (W1.2) where $a$ is the lattice constant, enabling a higher optical quality factor, compared with a W1 design for fixed slot widths[25]. The W1.2 slot cavity resonance also has less dependence on slot width compared to the W1 slot cavity resonance, important to improve the deep-UV (DUV) nanofabrication tolerances. While optical quality factor is higher for narrower slots and 80 nm slots have been fabricated with electron beam lithography, in this work we designed and worked with 120 nm slot widths to be compatible with the current design rule of our CMOS photolithography processes.

To achieve the integrated deeply-sub-wavelength PhC slot and the epitaxial active Ge detectors simultaneously, we developed the nanofabrication process flow principle to start first with the 120 nm slot optomechanical oscillator definition, followed subsequently by the monolithic *p-i-n* Ge detector epitaxy, vias, and electrode contact pads definition, and later by the input/output coupler fabrication and PhC nanomembrane release. The integration consists of 20 multi-level masks alignments and about 280 optimized nanofabrication process steps, across the 8" wafer sets. Figures 1b to 1d illustrate the nanofabricated slot cavities with high yield. We note that, first, a 100 Å oxide is deposited on pristine silicon-on-insulator wafers to: (1) achieve the 120 nm slots in a 248-nm DUV lithography stepper, (2) protect the Si surface for subsequent epitaxy growth on a clean Si lattice, and (3) protect the patterned PhC surface during the Ge detector process steps. This is followed by a *p+* implantation to define the bottom contact of the Ge detectors.

For the deeply-subwavelength slots, the patterned resist profile is rigorously numerically modeled and optimized for a 185 nm slot line width, which is then tightly process controlled with sloped oxide etching to transfer into a 120 nm slot in the silicon devices as shown in Figure 1c and 1d. All the PhC cavities, lattices, and subsequent process steps are aligned across the wafer. Next a monolithic Ge layer is epitaxially grown as described in earlier studies[29,30], followed by top *n+* implantation, vias definition, and metallization steps (as shown in Figure 1e). To maintain planar surfaces in the complete process, four planarization steps are introduced and interspaced across the entire process flow, involving oxide backfilling and multiple chemical-mechanical polishing with ~ 200 Å (initial levels) to ~ 1000 Å (latter levels) thickness variations in the multilayers across the wafer. Subsequently the input/output couplers are defined with an oxide over-cladded coupling waveguide (as shown in Figure 1f) and silicon inverse tapers, for input/output coupling loss less than 3-dB per facet.

Figure 2a shows the DC I-V diode characterization for the vertical *p-i-n* detector. The measured dark current is 500 nA at -1 V bias for our 4 μm × 25 μm Ge detectors while a dark current of 1 μA is the typical upper bound for our high-bandwidth detectors[29,30]. The measured 3-dB bandwidth of the detector is 9 GHz at 0 V bias and 18.5 GHz at -1 V bias as shown in Figure 2b, which agrees with our theoretical estimates detailed in Supplementary Information II. The oscillator-integrated detector responsivity is measured as 0.58 A/W at 0 V and 0.62 A/W near -0.5 V under 1550 nm illumination of 200 μW. With the integrated on-chip detector-oscillator, the measured optical signal-to-noise is ~ 10 dB from the spectrum analyzer measurement and the detector noise-equivalent-power is determined to be ~ 16 pW/$\sqrt{Hz}$.

Figures 2c and 2d illustrate examples of the measured optical transmission spectra of the slot cavity resonances (see Methods and Supplementary Information I). Two-mode resonances are observed in the transmission, corresponding to the fundamental (~ 1541.5 nm ) and higher-order mode excitations in the photonic band gap[12,26], and with typically loaded optical quality factors in the range of 60,000 to 150,000 for the fundamental mode (estimates of intrinsic quality factor are detailed in Supplementary Information III). The higher-order mode shows loaded quality factor typically in the range of 20,000 to 100,000. The modeled $|E|^2$ field distributions of the two resonances are shown in Figure 2c inset, with intrinsic quality factor of ~ 800,000 for the fundamental mode.

Figures 2e and 2f show the measured RF spectra of the integrated optomechanical oscillator at blue detuning and below/above the threshold power, respectively. The fundamental in-plane mechanical mode induced by radiation pressure is detected between 110 MHz to 120 MHz, depending on different sizes of the air-bridge (rectangular) holes in our design. As an example, for a dropped-in power about -15 dBm, the measured RF spectra for both detectors in room temperature and atmosphere show the fundamental mechanical resonance at 110.3 MHz and a cold cavity mechanical quality factor $Q_m$ of about 480. The modeled modal resonance displacement field is shown in the inset. By comparing results from external detector and integrated detector simultaneously, we note that our integrated Ge detector has low background noise floor that can go down to approximately -98 dBm (Figure 2e). The integrated oscillator-detector chipset, however, has a lower signal-to-noise ratio (SNR) due to the current short length (and effective length) of the Ge detector, which give ~ 50% absorption while on the other hand ensures a higher frequency response bandwidth. Excess noise spikes arise from the measurement background from the low signal when electrically read out from the chip. When driven (~ 400 µW) above threshold (~ 127 µW in this example), the intrinsic mechanical energy dissipation is overcome and the optomechanical resonator becomes a self-sustained OMO with narrow linewidths (~11 Hz in this example)[31], as illustrated in Figure 2f. We note that in Figure 2f the RF spectrum from the integrated Ge detector is amplified so that the output signals are at the same power levels for comparison. The vacuum optomechanical coupling rate is determined experimentally by introducing phase modulation on the input laser and comparing the peak density power for modulation frequency and mechanical frequency[23,32]. The vacuum optomechanical coupling rate is determined to be ~ 800 kHz which is much larger than other non-PhC optomechanical cavities[20], important to reduce the OMO threshold power and improve the transmitted SNR. The discrepancy of optomechanical coupling strength in simulation and experiment is due to coupling to other flexural modes[12,28]. Pertinent details of the oscillation threshold, optomechanical coupling rate, and loss channels are detailed in Supplementary Information IV.

Figure 3a shows the high-order harmonics (captured by a 12 GHz external photodetector) from our monolithic OMO-detector chipset with increased pump power, due to the nonlinear optomechanical transduction from the optical lineshape. With increased dropped-in power up to 3.2 mW (25 times of the threshold power, see Supplementary Information IV), we observed RF harmonics up to 6.9 GHz, the 59th harmonic in this device case, which is bounded by the spectrum analyzer measurement range. Such high-harmonics can serve as high frequency reference. While the linewidth of the high-harmonic modes can be broadened, harmonic-locking schemes can also be introduced to stabilize the entire OMO frequency spectra[33,34]. The higher-order harmonics can also be locked to optical transitions of atomic clock to improve the OMO long-term stability[24]. As an example, we demonstrated the injection locking of OMO by introducing phase modulation of the input laser at frequency close to OMO's fundamental frequency as shown in Figure 3b and 3c. By sweeping external modulation frequency towards OMO fundamental frequency, we observed the transitions from frequency pulling/mixing to quasi-locking, and then to fully-locked regimes. When the OMO fundamental frequency is locked to the external modulation drive, the high-order harmonics are also stabilized and have very narrow linewidth, as shown in Figure 3c for the 31st harmonics at 3.63 GHz. Moreover, we also observe the cooperative interaction between the OMO displacement, free carrier density, and temperature in a single device which leads to the generation of rich subharmonics[35]. As shown in Figure 3d, under various laser-cavity detunings and dropped-in powers, we can selectively excite one-half, one-third, and one-quarter subharmonic frequencies and their respective high-order harmonics. With the generation of both harmonics and subharmonics, our OMO device can be tuned to function both as a frequency multiplier and also a frequency divider in a single optomechanical cavity.

For the RF reference applications of OMOs, phase noise is an important character of a self-sustained oscillator. The phase noise performance of a OMO has been examined theoretically[14] and experimentally[2,11] in prior studies. Figure 4a shows the single-sideband phase noise spectra of our free-running OMO chipset, for a 112.7 MHz carrier (see Methods). In room temperature and atmospheric non-vacuum, our integrated OMO chipset exhibits a phase noise of approximately -103 dBc/Hz at 1 kHz offset and -125 dBc/Hz at 10 kHz offset, one of the lowest noise to date in reported OMOs[2,11,24]. For a comparison, phase noise measurements from the on-chip Ge detector and external detector are presented in Figure 4a. The integrated Ge detector exhibits lower phase noise at close-to-carrier offset (100 Hz to 10 kHz) and relatively higher phase noise at far-from-carrier offset (10 kHz to 10 MHz). We note that at higher frequency offsets (such as 1 MHz or more), the noise floor is limited only by our detector currently as the phase noise measured simultaneously by external detector can get as low as -165 dBc/Hz at 10 MHz offset. The higher phase noise at far-from-carrier offsets for the integrated Ge detector is a direct result of low SNR and large white noise floor from the RF amplification as indicated in the RF spectrum of Figure 2f. Figure 4a also plots the phase noise of the commercial electrical RF signal generator (Stanford Research System, Model SG384, DC-4.5 GHz) for comparison. As we can see, for offsets close-to-carrier frequency, our free running OMO has very significant amount of $1/f^3$ whereas for offsets far-from-carrier frequency ($f$ >100 kHz), our OMO actually has a lower phase noise performance.

The free-running OMO phase noise can be described by a closed-loop Leeson model[2,13,36] and consists a dependence of $1/f^3$ between 100 Hz and 1 kHz and a dependence of $1/f^2$ between 1 kHz and 10 kHz. From Leeson model, the $1/f^3$ and $1/f^2$ phase noise are due to $1/f$ flicker noise and $1/f^{\,0}$ white noise in the system. The Leeson frequency and corner frequency then obtained through a power-law fit of the phase noise plot as $f_L$ = 3 kHz and $f_c$ = 20 kHz respectively (the theoretical power-law model[36] is detailed in Supplementary Information V). Note that the measured $f_L$ is much larger than the oscillation linewidth (~11 Hz, see Figure 2f) measured by spectrum analyzer. This indicates our system has excess $1/f$ flicker noise component at low frequency offset which comes from slow environmental fluctuation such temperature or stage position shift. This explains the phase noise measured by integrated Ge detector has a lower phase noise in the close-to-carrier offset , since it is integrated in the same chip and less sensitive to drifts in stage positioning and  optical coupling(see Supplementary Information V for more information). We also note that, for both phase noise curves, there are further drops in phase noise level after the $1/f^{\,0}$ white phase noise as shown in the 1 to 10 MHz offset in Figure 4a. The phase noise behavior is beyond the classic Leeson model and is contributed from the pump laser phase noise, as theoretically predicted in Ref. 14.

The $1/f^{\,3}$ flicker frequency noise of the OMO can also be greatly reduced by introducing active or passive locking schemes with external master frequency references. We measured the phase noise of the OMO chipset under injection locking[33,34]. One distinct difference from the phase noise of free-running OMO is that the $1/f^{\,4}$ random walk frequency noise and $1/f^{\,3}$ flicker frequency noises are significantly suppressed below the 1 kHz offset. The reference phase noise shown in Figure 4b are measured by tuning the amplitude-modulated laser wavelength far from resonance, and measuring with the on-chip Ge detector. External RF gain is used before phase noise analyzer to keep the reference signal with the same RF power as the injection locking measurement. Comparing the phase noise directly from RF signal generator, we also note that additional components such as the electro-optic modulator (EOM) can add white noise at high offset frequencies due to the frequency transfer from the electronic circuits to optical carrier as shown in Figure 4b. This again demonstrates the unique advantage of OMOs in optical-RF signal processing without requiring electronic intermediates.

The timing jitter of the oscillator is calculated from the measured phase noise (see Supplementary Information V). For our free-running OMO the root-mean-square timing jitter, integrating the phase noise from 100 Hz to the carrier frequency (112.7 MHz), is 3.42 ps for the integrated detector and 10.01 ps for the external photodetector, with performance close to commercial electronic frequency standards. Allan deviation is another time-domain metric to characterize the frequency reference stability, computed from the oscillator phase noise by[36]

$$\sigma(\tau) = \sqrt{\sigma^2(\tau)} = \sqrt{\int_0^\infty \frac{4f^2 L_\varphi(f)}{v_0^2} \frac{\sin^4(\pi f \tau)}{(\pi f \tau)^2} df} \ .$$

Here $L_\varphi(f)$ is the oscillator phase noise and $v_0$ is the carrier frequency. Figure 4c shows the open-loop Allan deviations calculated from raw phase noise and power-law fitted phase noise for the free-running OMO. The consistency between different methods can be seen in Figure 4c where there is small phase noise discrepancy at the close-to-carrier and far-from-carrier offsets. Figure 4d shows the Allan deviations under injection locking scheme which also illustrates the longer term of stability.

In summary, we illustrated a CMOS-compatible integrated RF oscillator chipset, where PhC optomechanical cavities with deeply-subwavelength slot widths are monolithically integrated with high-bandwidth epitaxial Ge *p-i-n* photodetectors. Optomechanical cavities with optical quality factor of ~ 100,000 are co-fabricated with high-yield across full wafers with DUV lithography and multiple planarization processes, for chip-scale integrated optomechanical oscillators. Our oscillator demonstrates a CMOS-integrated radiation-pressure-driven oscillator with high 59th harmonic up to 6.9 GHz and selectively excited subharmonic tones. For practical applications in frequency references, we demonstrated the single-sideband phase noise of -125 dBc/Hz at 10 kHz offset with 112.7 MHz carrier frequency in room temperature and atmosphere, one of the lowest phase noise optomechanical oscillators to date. The chip-monolithic Allan deviation is observed down to $5\times10^{-9}$ at 1-millisecond integration, also at ~ 400 μW dropped-in powers, and likewise in room temperature and atmosphere operating conditions. Our work presents a promising step towards fully on-chip applications of optomechanical oscillators in the optical-RF information processing architectures.

**Methods**

**Chipset nanofabrication:** The CMOS-compatible process consists of 20 masks and multi-level alignments and about 280 optimized nanofabrication process steps, on a 8" silicon wafer at the foundry. The designed process flow principle starts with definition of the 120 nm critical dimension slot widths in the optomechanical oscillator (on substrate, without membrane release), followed by the epitaxial and vertical *p-i-n* Ge photodetector growth and vias/electrode contact pads definition.

To achieve the deeply-subwavelength slots on a 248-nm DUV lithography stepper, the resist profile is patterned with a 185 nm slot line width, which is then transferred into the oxide, with a residual slope in the oxide etch. The bottom 120 nm oxide gap is then etched into the silicon device layer through tight process control in the silicon etch, with resulting cavities shown in Figure 1c and 1d. Next all the PhC cavity and lattice patterns are aligned to the slot arrays across the 8" wafer and optimally etched into the device layer, to create the optomechanical cavity (unreleased). *p+* implantation for the bottom contact is implemented before the 500 nm Ge epitaxial growth for the detector. Vertical vias are next patterned and metal layer deposited for the contact holes and the patterned electrode pads as shown earlier in Figure 1e. Multiple planarization steps are interspaced across the entire process flow, critical for the success and high-yield of the OMO chipset. After fabricating the integrated Ge detectors, the optical input/output couplers are defined, with silicon inverse tapers and oxide over-cladded coupler waveguides with coupling loss less than 3-dB per facet. The silicon taper and oxide over-cladded waveguide has a modelled 98% coupling efficiency. The PhC cavity is next carefully released by etching the bottom oxide with buffered-oxide etch and tight process control.

**Device design:** The device design consists three steps. In the first step we simulate and optimize optical performance of the device. The photonic crystal band structure is simulated by MPB mode solver[36]. The field distribution and the quality factor of the slot-guided mode are simulated by free finite-difference-time-domain (FDTD) based solver MEEP[37] and finite-element-method (FEM) based commercial software COMSOL, for a comparison. Then we simulate the mechanical modes' displacement field and quality factor with the structure mechanics module of COMSOL. Finally, both optical and mechanical simulation result is combined in MATLAB to calculate the optomechanical coupling rate.

**RF and optical measurements:** The drive tunable diode laser is the Santec TSL-510C, tunable from 1510 to 1630 nm. A fiber polarization controller and a polarizer is used to select the transverse-electric (TE) state-of-polarization to drive the optomechanical oscillator. The external photodetector is a New Focus 125 MHz detector used to monitor the RF spectra, along with a slow detector to simultaneously track the optical transmission. When an optical amplifier is used, an isolator is included to protect the amplifier against potential damage from large optical reflections. For the RF spectrum measurements of integrated Ge detector, a RF probe (Picoprobe GSG-100-P, GGB Industries, Inc.) is contacted onto the aluminum pads.

**Phase noise measurements:** The output signals of both integrated Ge detector and external photodetector are amplified by low noise amplifiers to reach the typical power requirements of the signal source analyzer (Agilent 5052A). Different RF amplification (between the detector and the signal source analyzer) is used for the integrated and external detectors so that the photodetector power is kept the same for proper phase noise comparison. The amplified signals are passed through a narrow-band filter to suppress the subharmonics and/or high-order harmonics. In the measurements, to identify the real phase noise of our OMO, the IF gain and correlation of the Agilent instrument for phase noise measurement are set as 50 dB and above 128, respectively. For injection locking measurement, an external RF reference modulates the laser by using an EOM (JDS Uniphase, OC192 10 Gb/s Amplitude Modulator) before coupling into the chipset.

**Acknowledgements**

The authors acknowledge discussions with Jia-Gui Wu, and the lightwave component analyzer loan from Keren Bergman. This work is supported by the Defense Advanced Research Projects Agency (DARPA) DSO optomechanics program with program managers Drs. R. Lutwak and J. R. Abo-Shaeer under contract FA9550-10-1-0297 | C11L10831.

**Author contributions**

X.L, Y.H., Y.L., J.F.M., S.W.H., T.G., A.H., and D.A.H. performed the measurements, J.Z, J.F.M., and P.C.H., and C.W.W. performed the designs and layouts, M.Y., D.W., G.Q.L., and D.L.K provided the nanofabrication and samples, and X.L., Y.H, G. W., and C.W.W. put together the manuscript with contributions from all authors.

**Additional information**

The authors declare no competing financial interests. Supplementary information accompanies this paper online. Reprints and permission information is available online at

http://www.nature.com/reprints/. Correspondence and requests for materials should be addressed to C.W.W.

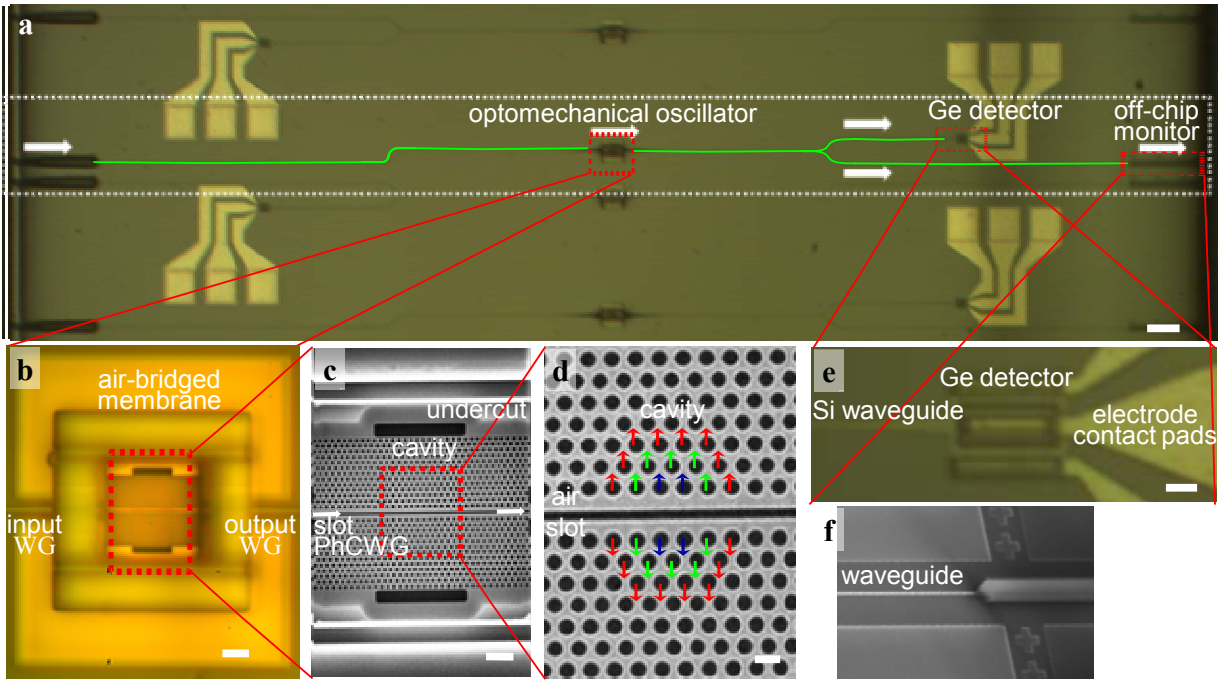

**Figure 1 | An integrated optomechanical oscillator chipset. a,** Optical image of designed integrated optomechanical oscillator (OMO). The dashed white box highlights the single device set, with the waveguide light paths shown in green. Drive laser is from the left, with two detection ports – an integrated monolithic Ge detector and an external monitor. Scale bar: 100 μm. **b,** Zoom-in optical image of designed optomechanical oscillator with in-line input/output waveguides (WG). Scale bar: 5 μm. **c,** Zoom-in scanning electron micrograph (SEM) of air-bridged photonic crystal slot cavity, along with optimized design input/output slot waveguides. The lattice constant $a$ is 500 nm and the ratio between hole radius and lattice constant $a$ is 0.34, centering the optical resonance within the photonic band gap and at 1550 nm. Scale bar: 2.5 μm. **d,** Zoom-in SEM of the slot cavity, formed by differential perturbative shifting of the nearest neighbor holes from a periodic lattice and denoted by the arrows (red: 5 nm; green: 10 nm; blue: 15 nm). Scale bar: 500 nm. **e**, Zoom-in optical image of designed Ge detector with tapered silica waveguide (left) and tapered electrode contact pads (right). Scale bar: 10 μm. **f,** Isometric view SEM of input silica waveguide with buried silicon inverse taper, for better impedance matching from fiber into the silicon waveguide. Scale bar: 500 nm.

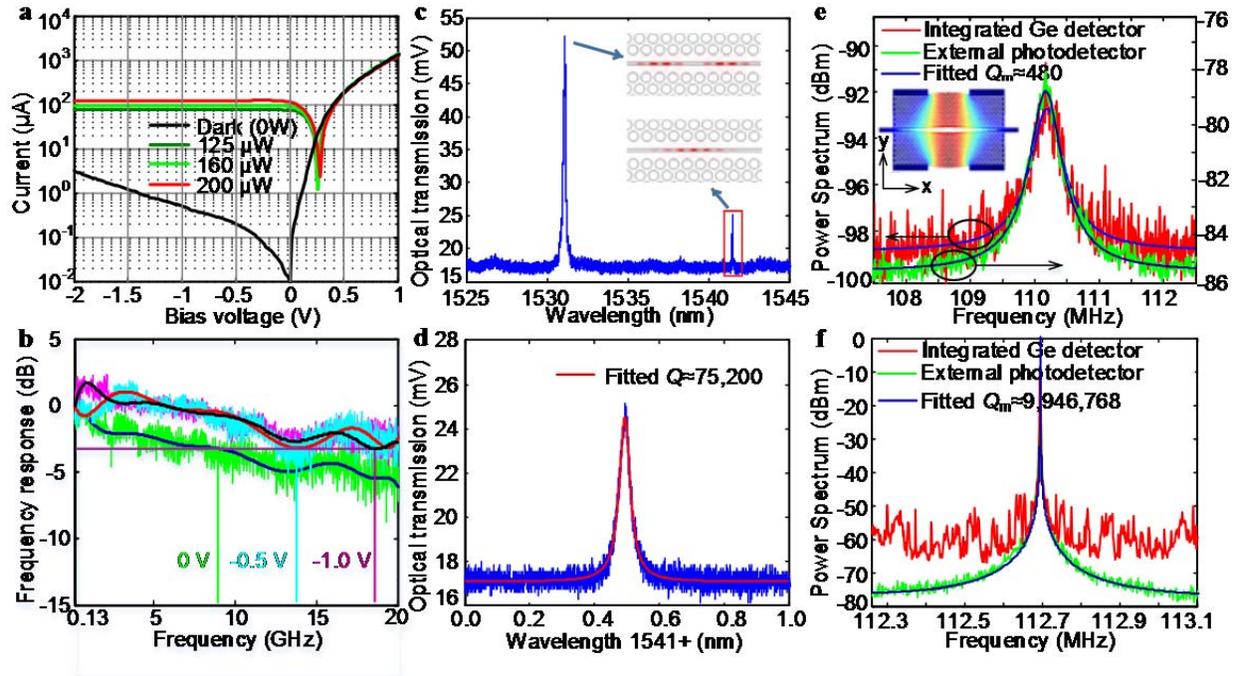

**Figure 2 | Photoresponse and optical/mechanical transmission of the monolithic detectors and optomechanical cavity. a,** Measured DC I-V curve for integrated Ge detector under dark and illumination conditions with different laser powers. **b,** Integrated Ge detector bandwidth under different reverse biases. Black (thicker) lines are the 9th degree polynomial fit for each bias. **c,** Transmission spectra with the two-mode resonances of the slot cavity. Inset: $|E|^2$-field distribution of the fundamental and higher-order resonances. **d,** Zoom-in of the fundamental (longer wavelength, and boxed in panel **c**) resonance with loaded $Q$ at 75,200. **e,** RF spectra with integrated Ge detector and external photodetector, of cold cavity regime before oscillation. Inset: Finite-element model of the fundamental eigenmode. **f,** RF spectra with integrated Ge detector and external photodetector of another device which shows mechanical mode centered near 112.7 MHz, under larger input laser power above threshold for oscillation mode. The output signals of both integrated Ge detector and external photodetector are amplified (~40 dB and ~10 dB, respectively) by low noise amplifiers to reach the typical power requirements of the signal source analyzer (Agilent 5052A).

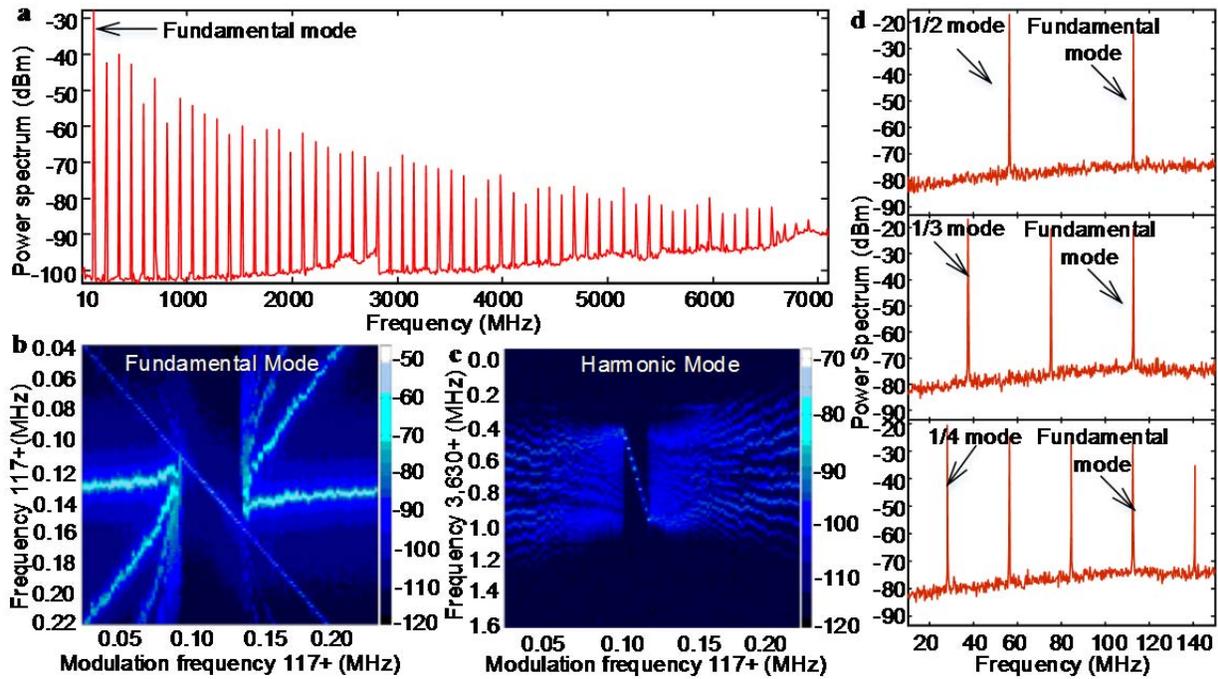

**Figure 3 | Integrated RF and harmonics measurements of monolithic radiation-pressure-driven optomechanical oscillator. a,** RF harmonics above threshold of the oscillator, up to the 59th harmonic at 6.9 GHz, with dropped-in power 3.2 mW. A 12 GHz photodetector (New Focus Model 1544) was used here to capture the harmonics. The resolution bandwidth (RBW) is 1 kHz and the video bandwidth (VBW) is 100 kHz. **b,** 2D spectra of the OMO signal with injection locking characteristics, with the horizontal axis as the tuned modulation frequency and the vertical axis as the RF spectra. **c,** 2D spectra of the high-order 31st harmonic at 3.63 GHz for the OMO signal with injection locking characteristics. **d,** RF subharmonics of the oscillator for (from top to bottom) half mode, one third mode, and quarter mode.

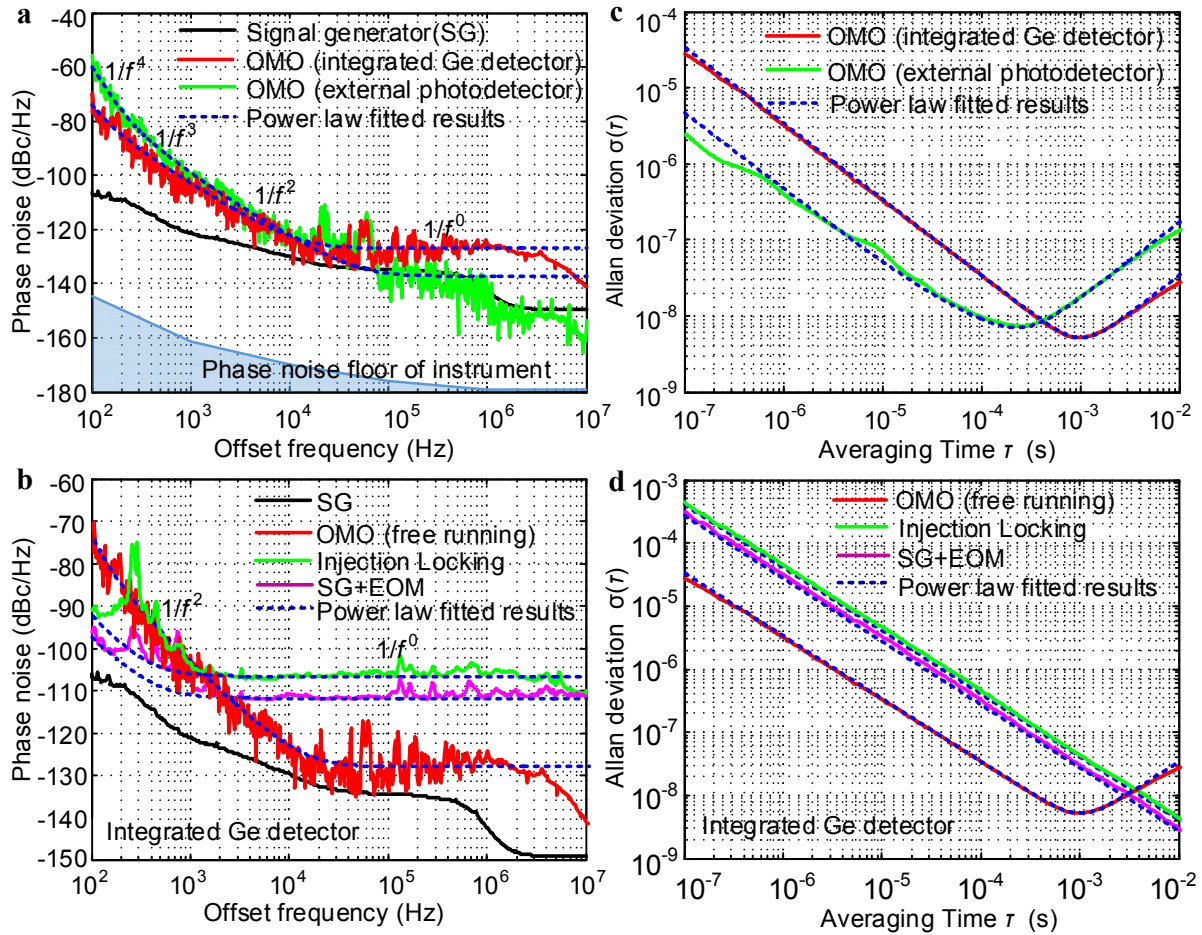

**Figure 4 | Phase noise and Allan deviation results. a,** Phase noise results for both integrated Ge detector and external photodetector at 400 μW dropped-in power. The phase noise of signal generator and the phase noise floor of the Agilent instrument are also shown in this panel for comparisons. The two blue dashed curves are the power-law fitted phase noise for the two detectors. **b,** Phase noise results for the injection locking and a signal generator modulated laser at off-resonance mode wavelength. The Phase noise of signal generator and the free-running OMO are also shown in this panel for comparisons. The three blue dashed curves are the power-law fitted phase noise. Note a slight noise peak between 200 Hz to 1 kHz in this free-running OMO data. **c,** The corresponding Allen deviation results converted from the measured phase noise results for both detectors in panel **a**. The blue dashed lines are the corresponding power low fitted results. **d,** The corresponding Allen deviation results converted from the measured phase noise results in panel **b**. The blue dashed lines are the corresponding power low fitted results.

# Supplementary Information for

# An integrated low phase noise radiation-pressure-driven optomechanical oscillator chipset

Xingsheng Luan,[1,*] Yongjun Huang,[1,2,*] Ying Li,[1] James F. McMillan,[1] Jiangjun Zheng[1], Shu-Wei Huang[1], Pin-Chun Hsieh[1], Tingyi Gu[1], Di Wang,[1] Archita Hati,[3] David A. Howe,[3] Guangjun Wen[2], Mingbin Yu[4], Guoqiang Lo[4], Dim-Lee Kwong[4], and Chee Wei Wong,[1,*]

[1]*Optical Nanostructures Laboratory, Columbia University, New York, NY 10027, USA*

[2]*Key Laboratory of Broadband Optical Fiber Transmission & Communication Networks, School of Communication and Information Engineering*

*University of Electronic Science and Technology of China, Chengdu, 611731, China*

[3]*National Institute of Standards and Technology, Boulder, CO 80303, USA*

[4]*The Institute of Microelectronics, 11 Science Park Road, Singapore 117685, Singapore*

*Author e-mail address: xl2354@columbia.edu, yh2663@columbia.edu; cww2104@columbia.edu

## I. Chip layout and experiment setup

The chip layout is shown in Figure S1. For compactness on the CMOS chip, we placed our integrated OMO as center symmetric pairs. Here the inverse oxide taper coupler is shown in red on both input/output coupling sides and the silicon waveguide is shown in blue. The Ge detector is shown in dark purple, with the electrodes, vias and contact pads labelled in purple outline. The TE polarized light (designed at 1550 nm) is first focused into a low-loss oxide coupler by free-space lens and then coupled into the silicon waveguide before reaching the slot-type PhC cavity. In order to achieve maximum coupling from oxide coupler to the OMO, a taper is introduced at silicon waveguide to slot PhC waveguide interface. For the coupling between waveguide and cavity, different designs are included such as direct tunneling coupling as shown in Figure 1c in main text and side-coupling shown in Figure S1d below. The transmitted light is then split into two paths; one into the integrated Ge detector and the other coupled out from the inverse oxide coupler to a free-space lens for external detector monitoring.

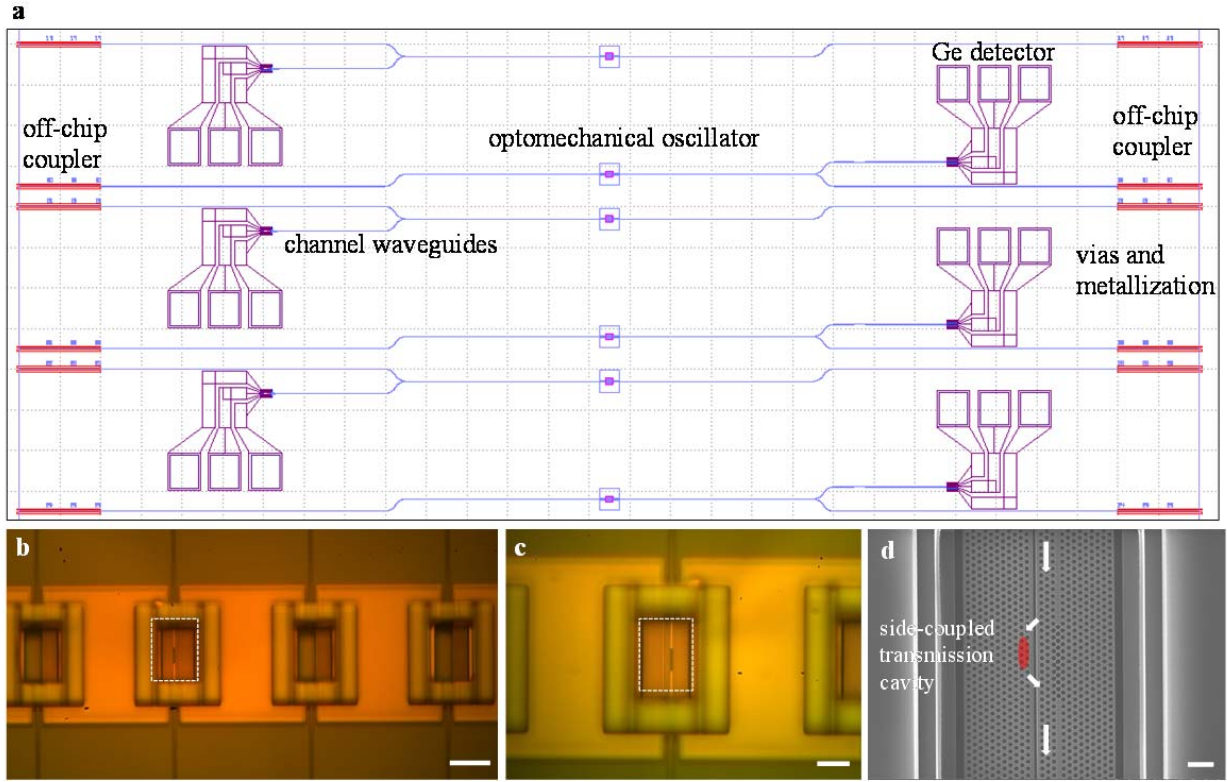

**Figure S1 | Layout and nanofabrication results of the integrated OMO chipset. a,** Overview layout within chipset. **b,** Optical image of nanofabricated OMOs. Scale bar: 30 μm. **c,** Optical image of nanofabricated OMO. Scale bar: 15 μm. **d,** SEM of OMO variations, with side-coupled tunneling transmission. Scale bar: 2 μm.

A simplified experimental setup is shown in Figure S2. Here the tunable laser (Santec TSL 510, Type C, 1500-1630 nm) first goes through a C-band erbium-doped fiber amplifier (EDFA), to achieve optical amplification if needed, such as for observing the higher order harmonics. For typical OMO operations, the EDFA is removed from the setup. Since our PhC cavity is designed for TE polarization, the fiber polarization controller and a bulk polarizer is used to eliminate the TM component of the input laser. At the output transmission, we use a slow detector (Thorlabs PDA10CS InGaAs Amplified Detector, bandwidth 17 MHz) and a fast detector (New Focus Model 1811 Low Noise Photoreceiver, bandwidth 125 MHz) to monitor the optical power and mechanical modulation on output field respectively. The electrical signal from Ge detector is measured from a RF probe (Picoprobe GSG-100-P, GGB Industries, Inc.) on the aluminum pads, as shown in Figure S2. To characterize the optomechanical coupling rate, an electro-optic phase modulator (EOM; Covega Mach-10 10G phase modulator) is added to generate phase modulation on the input light. The external frequency reference is a tunable signal generator (Stanford Research System, Model SG384, DC-4.5 GHz).

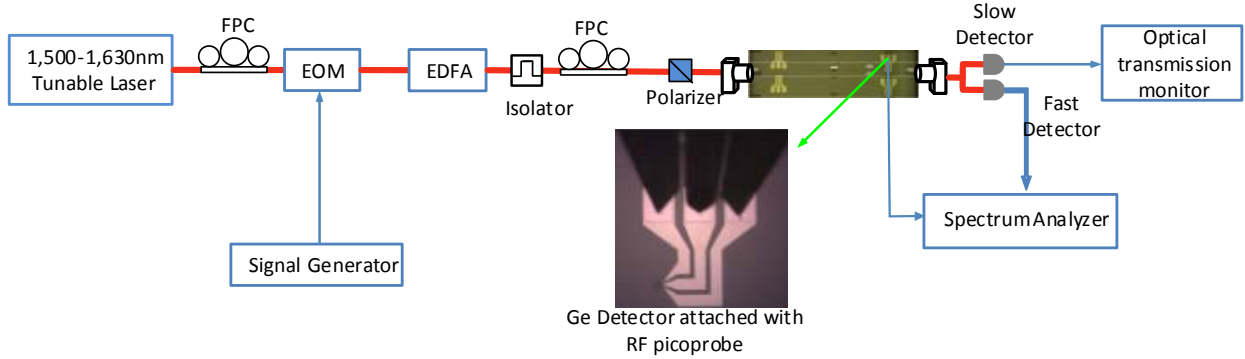

**Figure S2 | Simplified experimental setup for the optical transmission and mechanical resonance measurements of the integrated OMO.** A tunable semiconductor diode laser drives the OMO, with capabilities of external phase-modulation injection locking, optical amplification if needed, and simultaneous RF spectral analysis, phase noise analysis, and optical intensity transmission monitoring. The RF probes on the integrated Ge detector are illustrated in the optical image.

## II. Ge detector design and characterization

The Ge detector is in a vertical *p-i-n* configuration, as shown in Figure S3. Such vertical *p-i-n* configuration enables low leakage current [S1], important for increasing signal to noise ratio. The *p+* and *n+* junctions are formed on Si and Ge regions respectively and are separated by an intrinsic Ge absorbing layer with thickness 500 nm. The vertical *p-i-n* has a width 4 μm and length 25 μm. Previous optical simulation shows such dimensions can efficiently absorb more than 80% of the incident light traveling in the waveguide [S1].

To fully characterize the performance of Ge detector, we first measured its DC response, as shown in Figure 2a in main text, which indicates a typical diode characteristic. The dark current is measured to be 500 nA at -1 V bias, while 1 μA is typically considered as the upper limit for a high-bandwidth design [S1]. Responsivity is another important parameter that characterizes the detector efficiency. For the Ge detector, we measured the responsivity with varying bias voltage. At zero bias (photovoltaic mode), the responsivity is almost 0.58 A/W. Increasing the reverse

S-3

bias (photoconductive mode) increases the responsivity of the detector. At -1 V bias, the responsivity of detector is saturated and reaches a maximum value of 0.62 A/W.

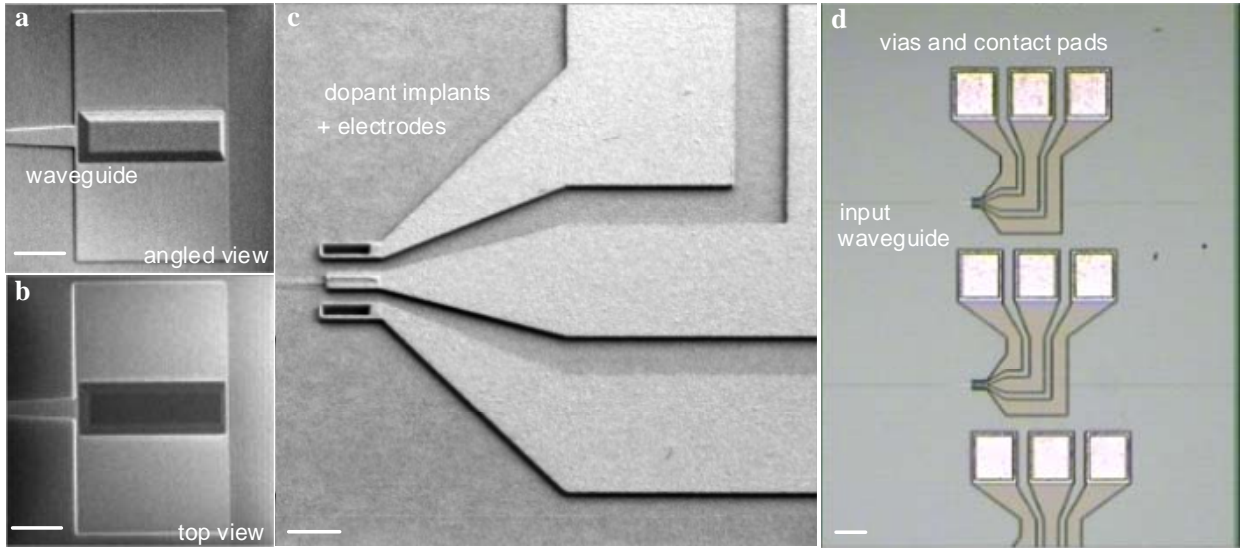

**Figure S3 | The SEM images of the integrated Ge detector. a**, Angled view. Scale bar: 10 μm **b**, Top view. Scale bar: 10 μm. **c**, Details of dopant implants and electrodes. Scale bar: 25 μm. **d**, Optical image of two Ge detectors with 25 μm and 50 μm lengths. Scale bar: 100 μm.

The frequency response of Ge detector is characterized by a lightwave component analyzer (LCA; Agilent 8703A, 1550nm, 0.13 to 20GHz) with the setup shown in Figure S4. Here the laser source from the LCA is modulated by a built-in modulator which is synchronized with its electrical measurement component. The frequency response of Ge detector under different bias voltage is shown in Figure 2b of the main text, a 9 GHz bandwidth at zero bias and 18.5 GHz at -1 V bias. Theoretically, the bandwidth is determined by

$$f_{3dB} = \sqrt{\frac{1}{1/f_{transit}^2 + 1/f_{RC}^2}}, \quad \text{(S-1)}$$

where

$$f_{transit} = \frac{0.45 v_{sat}}{t_{i-Ge}} \text{ and } f_{RC} = \frac{1}{2\pi RC} \quad \text{(S-2)}$$

are the frequencies from the carrier transit time under saturation and the cutoff frequency of the capacitor formed by *p-i-n* junction, respectively. At zero bias, calculations indicate the bandwidth of the detector is limited by the cutoff frequency of the *p-i-n* junction at 20 GHz.

However, the experimental measurements are lower than this possibly due to the large lattice mismatch with Si (~4.2%) or variations in the actual RC time constant [S1].

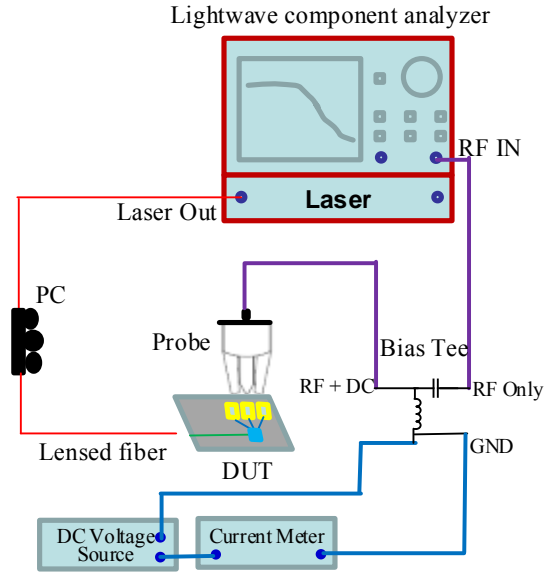

**Figure S4 | Setup for characterizing the DC I-V and RF bandwidth properties of the integrated Ge detector.** A RF probe contacts the metal pads as shown in Figure S2. Measurements are typically averaged over 10 scans, and the laser from lightwave component analyzer is synchronized with its built-in electronic measurement circuits to measure the RF response from on-chip Ge detector.

### III. Estimation of intrinsic quality factor

To determine the intrinsic quality factor, we first determine the ratio of coupling rate and total loss rate $\kappa_{ex}/\kappa$. From the coupled-mode theory of the intracavity field [S2]

$$\dot{\alpha} = -\frac{\kappa}{2}\alpha + i\Delta\alpha + \sqrt{\frac{\kappa_{ex}}{2}}\alpha_{in}, \tag{S-3}$$

where $\alpha$ is the normalized intra cavity field amplitude in the rotating frame with laser frequency $\omega_L$ and $\Delta = \omega_L - \omega_{cav}$ is the laser detuning with respect to the cavity mode. The factor 2 for $\kappa_{ex}$ is because our cavity is a bidirectional coupling standing-wave cavity [S2] and single direction is divided by 2. The steady state solution reads

$$\alpha = \frac{\sqrt{\frac{\kappa_{ex}}{2}}\alpha_{in}}{\frac{\kappa}{2} - i\Delta}. \tag{S-4}$$

Thus the transmitted light is given by

$$\alpha_{out} = i\sqrt{\frac{\kappa_{ex}}{2}}\alpha = \frac{i\frac{\kappa_{ex}}{2}\alpha_{in}}{\frac{\kappa}{2}-i\Delta}. \quad (S\text{-}5)$$

For the peak power,

$$\Delta = 0 \text{ and } \alpha_{out} = \frac{i\kappa_{ex}\alpha_{in}}{\kappa}. \quad (S\text{-}6)$$

Thus the output power and input power are related by

$$\frac{P_{out}}{P_{in}} = \frac{\hbar\omega_L |\alpha_{out}|^2}{\hbar\omega_L |\alpha_{in}|^2} = \left(\frac{\kappa_{ex}}{\kappa}\right)^2. \quad (S\text{-}7)$$

On the other hand, the intrinsic quality factor and loaded quality factor are also related by coupling rate and total loss rate

$$\frac{Q_L}{Q_i} = \frac{\omega_L/\kappa}{\omega_L/\kappa_i} = \frac{\kappa_i}{\kappa} = 1 - \frac{\kappa_{ex}}{\kappa}. \quad (S\text{-}8)$$

By inferring the input power before cavity and output power after cavity, we can determine the intrinsic quality factor. In this case, we also note that the accuracy is limited by the uncertainty of loss introduced by coupling from waveguide mode to slot guided mode due to effective index mismatching, which functions as an unknown attenuation before the cavity (and also after the cavity). To avoid this uncertainty, we infer $\kappa_{ex}/\kappa$ by comparing the peak power with nearby waveguide mode which lies out of the photonic crystal band gap. For waveguide mode outside the photonic crystal band gap, the photonic crystal slab functions as a multimode interference structure and, for an upper bound on the $Q_i$ estimate, we have $P_{out,max}/P_{in} \sim 1$ for the maximum transmitted wavelength. Thus, the ratio between cavity mode measured peak power and observed maximum transmitted power (as shown in Figure S5) equals the ratio $P_{out}/P_{in}$ for cavity mode, just like a side-coupled cavity.

From Figure S5, we get

$$\frac{P_{out}}{P_{in}} \approx \frac{0.02518}{0.05676} = 0.4436, \quad (S\text{-}9)$$

so

$$Q_i = Q_L\left(1-\frac{\kappa_{ex}}{\kappa}\right)^{-1} = 213,195. \quad (S\text{-}10)$$

So the intrinsic quality factor is estimated to be ~ $2\times10^5$. Note that for the real case (even at maximum transmitted wavelength), $P_{\text{out,max}}/P_{\text{in}} < 1$, thus our above estimate is an upper bound for the intrinsic quality factor.

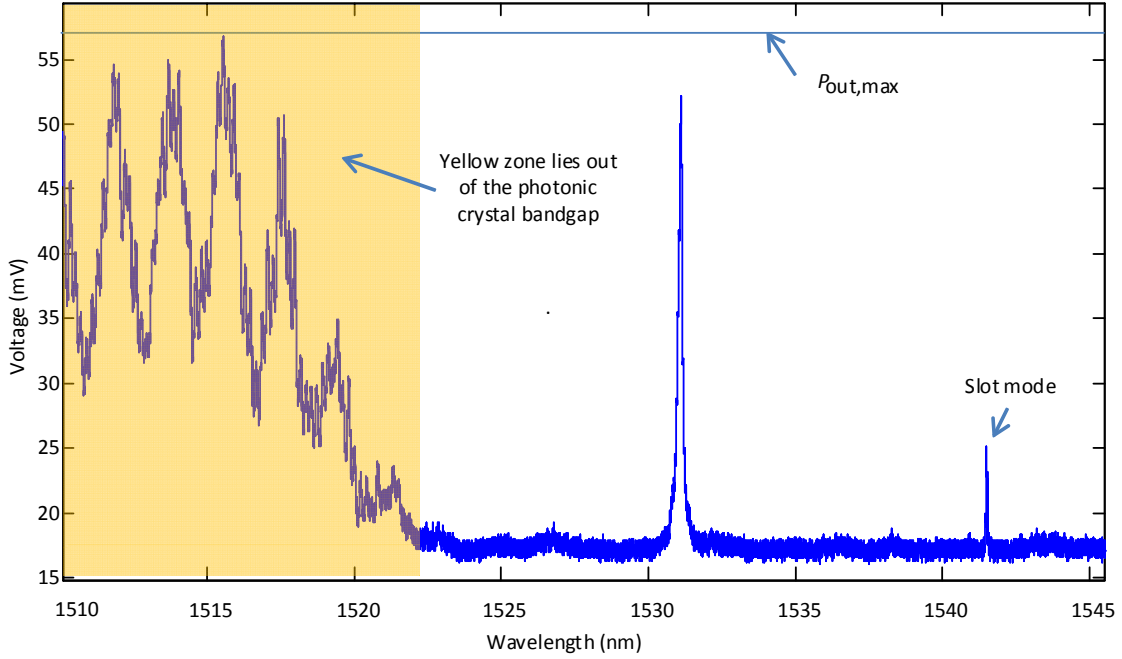

**Figure S5 | Optical transmission in wide wavelength for the estimation of intrinsic quality factor.** Wavelength in the yellow regime lies out of photonic bandgap and thus have larger transmission. The Fabry-Perot like transmission is due to internal reflections inside the waveguide. The level of maximum power and the slot mode is also identified in this plot.

## IV. Determination of threshold power, optomechanical coupling rate, and loss channel

Power consumption is another important factor for a frequency reference. For our fully integrated OMO, the operating power is slightly above 200 µW. This is determined by the threshold power of parametric oscillation of the optomechanical resonator, where the effective linewidth becomes zero. When effective linewidth is zero, the change of linewidth due to optomechanical interaction is given by [S3,S4]

$$\Gamma_{opt} = g_0^2 \bar{n}_{cav} \left( \frac{\kappa}{(\Delta+\Omega_m)^2 + (\kappa/2)^2} - \frac{\kappa}{(\Delta-\Omega_m)^2 + (\kappa/2)^2} \right), \quad \text{(S-11)}$$

where

$$\bar{n}_{cav} = \frac{P}{\hbar\omega_L} \frac{\kappa_{ex}}{\Delta^2 + (\kappa/2)^2} . \tag{S-12}$$

Let

$$\Gamma_{eff} = \Gamma_{opt} + \Gamma_m = 0, \tag{S-13}$$

we get

$$P_{th} = \frac{\Omega_m}{Q_m} \frac{\hbar\omega_L \left(\Delta^2 + (\kappa/2)^2\right)}{g_0^2 \kappa_{ex} \kappa} \left( \frac{1}{(\Delta - \Omega_m)^2 + (\kappa/2)^2} - \frac{1}{(\Delta + \Omega_m)^2 + (\kappa/2)^2} \right)^{-1}. \tag{S-14}$$

Under weak retardation approximation ($\kappa_{ex} \gg \Omega_m$), we can simplify the expression above

$$P_{th} = \frac{\hbar\omega_L}{4g_0^2 \kappa_{ex} \kappa Q_m} \frac{\left(\Delta^2 + (\kappa/2)^2\right)^3}{\Delta}. \tag{S-15}$$

Optimizing the detuning, the minimum threshold power is

$$P_{th} = \frac{27}{400\sqrt{5}} \frac{\hbar\omega_L}{g_0^2} \frac{\kappa^4}{Q_m \kappa_{ex}}, \tag{S-16}$$

and the corresponding detuning is

$$\Delta = \frac{\kappa}{2\sqrt{5}}. \tag{S-17}$$

Plugging experimental parameters into the expression above, we get $P_{th} \approx 127$ μW.

From Eq. (S-16), we get

$$P_{th} \propto \frac{1}{g_0^2 Q_m Q_o^3}. \tag{S-18}$$

This inverse dependence on mechanical $Q_m$, inverse-cubic dependence on the optical $Q_o$, and inverse-square dependence on vacuum optomechanical coupling rate make it desirable to have high mechanical/optical $Q$ and large optomechanical coupling rate, to achieve low operating powers. In measurements the threshold power is characterized by slowing varying the input power and monitoring the mechanical spectrum, as shown Figure S6a to S6c.

The vacuum optomechanical coupling strength is determined by phase modulating the input light and compare the peak density power at OMO frequency and external modulated frequency in the spectrum domain [S5,S6]

$$g_0^2 = \frac{1}{2\bar{n}_{th}} \frac{\phi_0^2 \Omega_{mod}^2}{2} \frac{S_{II}^{meas}(\Omega_m) \cdot \Gamma_m / 4}{S_{II}^{meas}(\Omega_{mod}) \cdot ENBW}. \tag{S-19}$$

Here, $\phi_0$ is the phase modulation amplitude and ENBW is abbreviated for effective noise bandwidth which is determined by the spectrum analyzer settings.

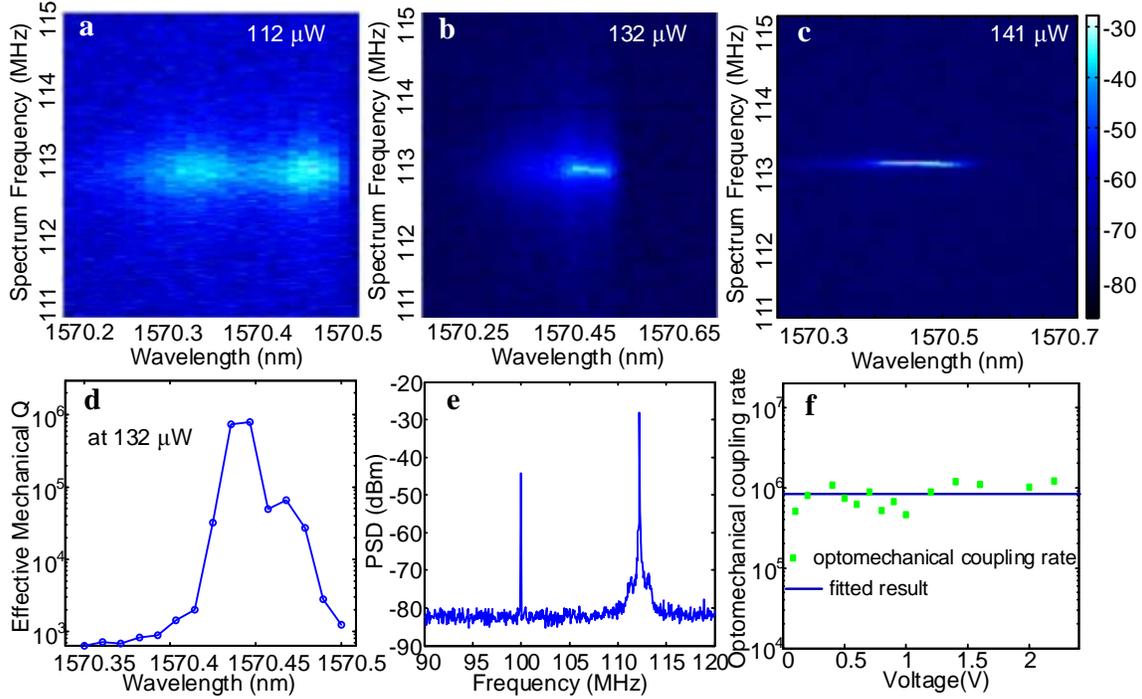

**Figure S6 | Measured power spectrum and calculated mechanical $Q$ and optomechanical coupling properties. a-c,** 2D power spectra versus detuning wavelength and RF frequency at different dropped-in power values. **d,** Derived mechanical $Q$ for the input power value shown in panel **b**. **e,** Power spectral density for the OMO signal and phase modulated signal. **f,** Calculated and fitted optomechanical coupling versus modulation voltage.

In experiment, we characterize the vacuum optomechanical coupling strength to be ~800 kHz (as shown in Figure S6e and S6f), which is smaller than numerical simulations of ~ 2.5 MHz. The discrepancy can be attributed to the deviations of the mechanical mode into two (slightly different) independent beams of the slot cavity and the mixing with other flexural modes of slot cavity. One example for the deviations is the splitting RF spectra obtained from out another chipset as shown in Figure S7.

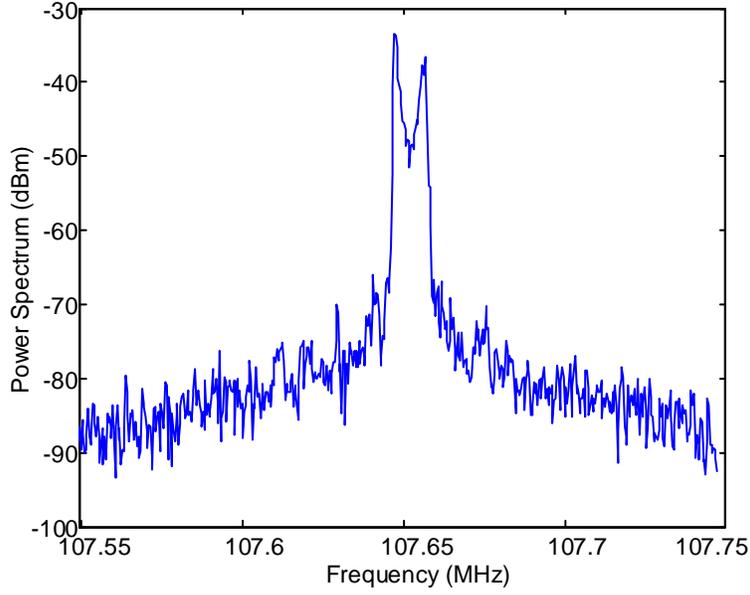

**Figure S7 | Frequency splitting due to the deviations of the mechanical mode into two (slightly different) independent beams of the slot cavity.**

In addition, several other reasons are responsible for the power consumption. First, observed under the top infrared imaging camera, ~ 5 dB of input light is scattered when coupling from the input silicon waveguide to slot cavity, mainly due to the finite tunneling rate between the waveguide mode and the cavity mode (highlighted in yellow in Figure S8). The finite tunneling rate of desired mode can also be verified in the optical transmission spectrum. As shown in Figure S5, the transmission of another higher mode has much larger power transmitted, which is proportional to the coupling rate $\kappa_{ex}$. This is because the higher order mode has a wider field distribution which leads to large field overlap and thus coupling rate with waveguide field. Moreover, a larger slot width leads to a lower effective modal index, with more scattering from the high-index silicon input channel waveguide due to mode mismatch. The coupling loss can be reduced by introducing more adiabatic coupling schemes into the slot waveguide and the side coupling approaches. Taking account the loss in coupling into the cavity, the dropped in power $P_d = \sqrt{P_{in}P_{out}}$ will be more useful in experimental analysis.

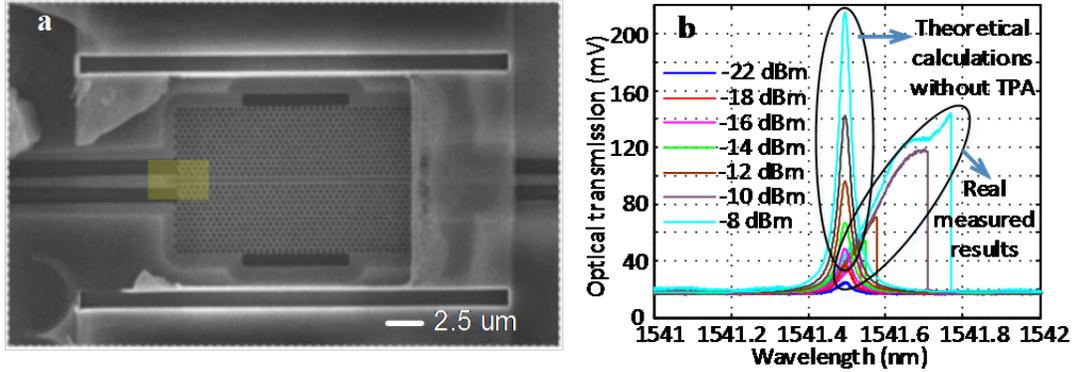

**Figure S8 | SEM view and measured/calculated optical transmission for the characteristics of loss channels. a,** SEM view of the OMO cavity. **b,** Measured versus calculated optical transmission.

When input power becomes large, nonlinear absorptions, e.g., two photon absorption and free carrier absorption, become an important loss channel [S6, S7]. As an example, we measured the thermal optical bistability effect by slowly sweeping the wavelength from shorter wavelength to longer wavelength (10 nm/s) under different input laser powers, as shown in Figure S8b. Theoretical ideal transmission resonances without nonlinear absorption is also shown in Figure S8b.

**V. Measurement of phase noise and theoretical model fits**

The phase noise of our OMO is measured by using Agilent 5052A signal source analyzer replacing the spectrum analyzer in the setup shown in Figure S1. To satisfy typical input power levels for the Agilent instrument, the output of both external photodetector and integrated Ge detector should be amplified by low noise amplifier. In measurement, to identify the real phase noise of our OMO, the IF gain and correlation of the Agilent instrument for phase noise measurement are set as 50 dB and above 128, respectively. At such settings, the phase noise floor of the instrument is shown in Figure 4a in main text, which is sufficient for our OMO measurement. The phase noise properties of three signal sources are measured for comparison, including a reference clock directly generated by a tunable RF signal generator (Stanford Research System, Model SG384, DC-4.5 GHz), and the OMO signal detected by both the external photodetector and integrated Ge detector. An example of the measured results are illustrated in Figure 4a of the main text.

To demonstrate the contributions of drifts in stage positioning and optical coupling on the phase noise for the integrated Ge and external detectors. We measure and compare the phase noise results at different conditions as shown in Figure S9 for examples. Specially, the lowest green curve is measured with laser modulated by EOM and then directly detected by the external detector while the dark blue curve is measured by first coupling the EOM-modulated laser into chip and then coupled out and detected by external detector. Here, the laser frequency is tuned far away from the optical cavity resonance and thus the mechanical oscillation is not amplified. By comparing these two groups of data, we can clearly see the difference is from the phase noise contributions of the drifts in stage positioning and optical coupling. Moreover, from the curves (dark blue and red, or light blue and purple), we can see lower phase noise in the close-to-carrier offset for the integrated Ge detector, which can confirm the drifts in stage positioning and optical coupling at output also contribute the phase noises. We can also compare the injection locking phase noise curves (light blue and purple) with the off-resonance mode curves (dark blue and red) to know that ~5 dBc/Hz phase noises is from OMO.

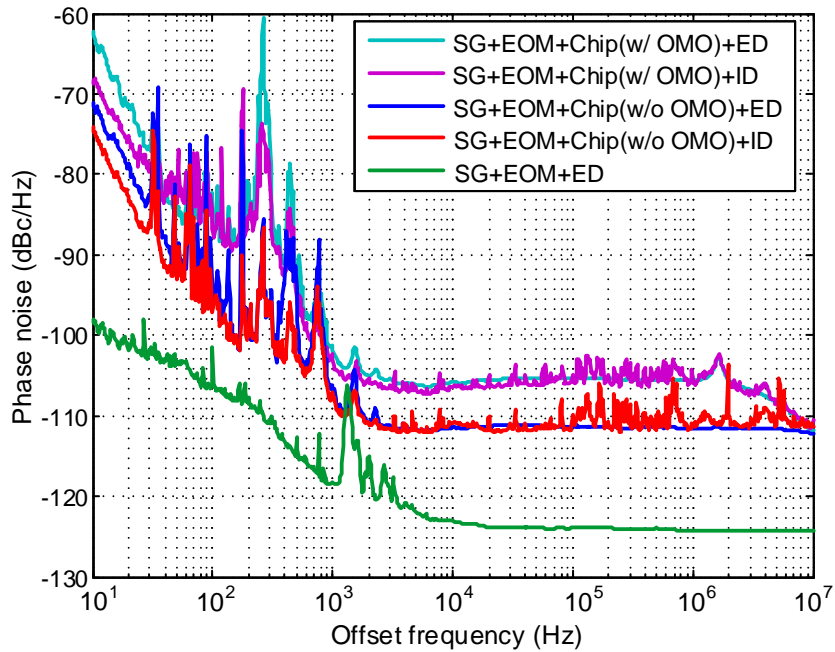

**Figure S9 | Phase noises for demonstrations of the contributions of drifts in stage positioning and optical coupling.** Here "ID" refer to integrated Ge detector and "ED" external detector.

Here we further theoretically fit the phase noise of OMO signal detected by the on-chip and off-chip detectors by using polynomial $\sum_{i=-4}^{0} b_i f^i$ (power law fitting theory) [S8]. Here $b_i$ is obtained directly from the measured phase noise level at the different frequency offsets, and the corresponding $b_i$ values obtained are: $b_0 = -127$ dBc/Hz, $b_2 = -103$ dBc/Hz, $b_3 = -90$ dBc/Hz, and $b_4 = -78$ dBc/Hz. As an example, the phase noise of our OMO with integrated Ge detector has a $1/f^4$ random walk frequency noise at lower offset related with mechanical shock, vibration, temperature, or other environmental effects, a $1/f^3$ flicker frequency noise in the range of 250 Hz to 800 Hz offset due to laser flicker phase noise, a $1/f^2$ white frequency noise in the range of 800 Hz to 15 kHz offset related to, and finally a $1/f^0$ white phase noise at higher frequency offsets. The $1/f^4$ noise is the result of slow environment noise processes, for example temperature fluctuation and instabilities from the measurement stage. This technical noise can be reduced by introducing temperature control and position feedback to the measurement stage, similar to the case for quartz oscillators. The $1/f^4$ may also arise from the diode laser used in our measurements. From Leeson model, the $1/f$ flicker phase noise, common in semiconductors, can convert into $1/f^3$ noise at low frequency offset in a closed-loop oscillator. Further characterization of the laser noise can be attained by comparing the cases where OMO is driven by a low-noise fiber laser instead of a semiconductor diode lasers which have inherent carrier relaxation dynamics.

Moreover, the root-mean-square (RMS) timing jitter can also be converted from the measured phase noise results,

$$\text{RMS } J_{PER}\big|_{f_1 \text{ to } f_c} = \frac{1}{2\pi f_c}\sqrt{2\int_{f_1}^{v_0} 10^{\frac{L_\varphi(f)}{10}}\,df}\ . \tag{S-20}$$

where $f_1$ and $v_0$ denote the start and stop (carrier frequency) of the integral. $L_\varphi(f)$ is the measured phase noise in dBc/Hz. We calculated several RMS timing jitter values by stating different frequency $f_1$ as shown in Table 1.

**Table 1 | Timing jitter for OMO measured by integrated Ge detector and external photodetector respectively.** Otherwise denoted, units are in ps and shows the phase jitter per frequency segment, from the label frequency to the carrier.

| Offset Frequency Range(Hz) | from 100 Hz | from 1 kHz | from 10 kHz |
|---|---|---|---|
| OMO (integrated Ge detector) | 3.42 | 2.56 | 2.53 |
| OMO (external photodetector) | 10.01 | 0.69 | 0.40 |

The Allan deviation is calculated from the measured phase noise by

$$\sigma(\tau) = \sqrt{\sigma^2(\tau)} = \sqrt{\int_0^\infty \frac{4f^2 L(f)}{f_c^2} \frac{\sin^4(\pi f \tau)}{(\pi f \tau)^2} df} \; . \tag{S-21}$$

Here $\sigma^2(\tau)$ is the Allan variance and $L$ is the phase noise of the oscillator. Furthermore, for better phase noise characteristics especially at close-to-carrier offset, injection locking technique can be used as a simple example to suppress the $1/f^4$ and $1/f^3$ noise. The setup is the same as Figure S1 and the electro-optic modulator used for injection locking is an amplitude modulator (JDS Uniphase, OC192 10 Gb/s Amplitude Modulator).

**Supplementary references:**